\documentclass[12pt]{article}

\usepackage{color,amsmath}
\usepackage{times}
\usepackage{graphicx}
\usepackage{multirow}

\usepackage[version=4]{mhchem}
\usepackage{comment}
\usepackage{hyperref}

\usepackage[sort,square,numbers]{natbib}
\usepackage{authblk}

\newcounter{lastnote}

\author{Seunghoon Lee}
\author{Huanchen Zhai}
\author{Garnet Kin-Lic Chan\thanks{gkc1000@gmail.com}}
\affil{Division of Chemistry and Chemical Engineering, California Institute of Technology, Pasadena, California 91125, USA}

\date{}

\begin{document} 

\baselineskip24pt

\title{An ab initio correction vector restricted active space approach to the L-edge XAS and 2p3d RIXS spectra of transition metal clusters}
\maketitle

\begin{abstract}
    We describe an  ab initio approach to simulate L-edge X-ray absorption  (XAS) and 2p3d resonant inelastic X-ray scattering (RIXS) spectroscopies.
    We model the strongly correlated electronic structure within a restricted active space and employ a correction vector formulation instead of sum-over-states expressions for the spectra, thus eliminating the need to calculate a large number of intermediate and final electronic states.
    We present benchmark simulations of the XAS and RIXS spectra of the iron complexes [FeCl${}_4$]${}^{-1/-2}$ and [Fe(SCH${}_3$)${}_4$]${}^{-1/-2}$ 
    and interpret the spectra by deconvolving the correction vectors. Our approach represents a step towards simulating the X-ray spectroscopies of larger metal cluster systems that play a pivotal role in biology.
\end{abstract}

\clearpage

\section{Introduction}

\begin{figure}[ht!]
\center{\includegraphics[width=0.7\columnwidth]{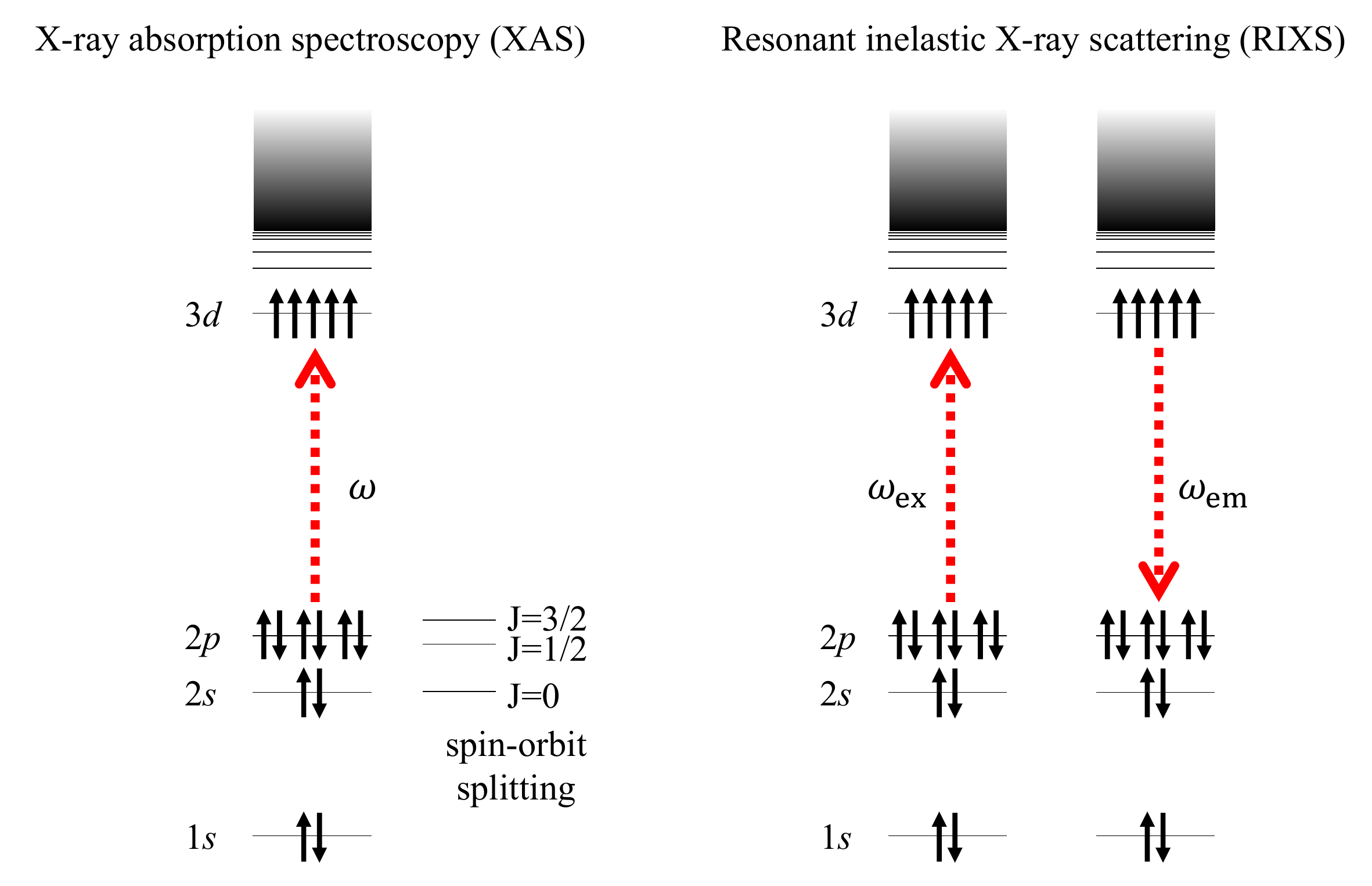}}
\caption{Schematic of the one- and two-step processes of L${}_{2,3}$-edge XAS and 2$p$3$d$ RIXS spectroscopies.}\label{fig:XAS_RIXS}
\end{figure}

X-ray spectroscopies are indispensable for characterizing the electronic structure of transition metal complexes~\cite{roemelt2013combined, kowalska2017iron, van2018electronic, maganas2019comparison}. 
For first-row transition metals,  L${}_{2,3}$-edge X-ray absorption  (XAS) and 2p3d resonant inelastic X-ray scattering (RIXS) spectroscopies are of particular interest~\cite{wasinger2003edge}.
They involve
one- and two-step transition processes, respectively, between the 2p-core and 3d-valence orbitals, is depicted in Fig.~\ref{fig:XAS_RIXS}. The dipole-allowed nature of these spectra ensures high intensity and energy resolution.
However, the effects of ligand-fields, presence of different multiplet spins, and spin-orbit coupling, complicate the interpretation of the spectra.
Theoretical models are thus essential. 

Recently, several ab initio methods have appeared to compute L-edge XAS~\cite{josefsson2012ab, roemelt2013combined, maganas2019comparison} and RIXS~\cite{josefsson2012ab} spectra.
These methods use sum-over-state expressions~\cite{josefsson2012ab, roemelt2013combined, maganas2019comparison}.
However, such approaches become impractical when there are a large number of intermediate and final states, as is expected to be the case when simulating larger bioinorganic clusters~\cite{kowalska2017iron, van2018electronic}.

A route to computing spectra is the correction vector (CV) formulation~\cite{soos1989valence}, where frequency-dependent response equations are solved to obtain the CVs, which determine  the spectrum at each frequency.
Here, we describe an ab initio implementation of the CV approach
for L-edge XAS and 2p3d RIXS spectra, within a restricted active space model of the correlated transition metal electronic structure~\cite{olsen1988determinant}.
The outline of the paper is as follows.
In section~\ref{sec:theory}, we introduce the ab initio relativistic Hamiltonian and the CV approach for L-edge XAS and 2p3d RIXS spectra.
We also describe how to deconvolve the spectra to separate different electronic effects. 
In section~\ref{sec:comput}, we describe  the geometries, molecular orbitals, active space models, wave function ansatz, and methods to optimize the wave function ansatz~\cite{chan2002highly, ronca2017time, zhai2022comparison}.
In section~\ref{sec:results}, we compute the XAS and RIXS spectra for monomeric ferrous and ferric tetrahedral iron complexes and compare them to available experimental spectra.
By deconvolving the theoretical spectra, we interpret the contribution of different electronic effects and states to the peaks in the XAS and RIXS spectra, highlighting the role of certain electron correlations. In section~\ref{sec:conclusions} we provide some perspective on future developments of this approach.

\section{Theoretical formulation}

\label{sec:theory}
\subsection{Spin-orbit Hamiltonian}

We start from an ab initio Hamiltonian containing spin-orbit coupling described within the mean-field Breit-Pauli (BP) approximation. In second quantization, this is
\begin{equation}
    \hat{H} = \sum_{ij}\sum_{\sigma=\alpha,\beta} h_{ij} \hat{a}^{\dagger}_{i\sigma} \hat{a}_{j\sigma} + \frac{1}{2} \sum_{ijkl} \sum_{\sigma,\tau=\alpha,\beta} (ij|kl) \hat{a}^{\dagger}_{i\sigma} \hat{a}^{\dagger}_{k\tau} \hat{a}_{l\tau} \hat{a}_{j\sigma} + \sum_{ij} \mathbf{h}^{\mathrm{BP}}_{ij} \cdot \hat{\mathbf{T}}_{ij},
\end{equation}
where $h_{ij}$ and $(ij|kl)$ are the ab initio one- and two-electron integrals,
\begin{align}
    h_{ij} & = -\frac{1}{2} (i| \hat{\nabla}^2 |j) + \sum_A (i| \frac{Z_A}{\hat{r}_{1A}} |j), \\
    (ij|kl) & = (ij| \frac{1}{\hat{r}_{12}} |kl),
\end{align}
 $\mathbf{h}^{\mathrm{BP}}_{ij}$ are the mean-field Breit-Pauli matrix elements, 
\begin{align}
    \mathbf{h}^{\mathrm{BP}}_{ij} & = ( i | \hat{\mathbf{h}}_{\rm en} | j )
    + \sum_{kl} D_{kl} \left(  (ij | \hat{\mathbf{h}}_{\rm ee} | kl)
    - \frac{3}{2} (il| \hat{\mathbf{h}}_{\rm ee} |kj) - \frac{3}{2} (kj| \hat{\mathbf{h}}_{\rm ee} |il) \right),\label{eq:one-e-int} \\
    \hat{\mathbf{h}}_{\rm en} & = \frac{\alpha^2}{2} \sum_{A} \frac{Z_A}{\hat{r}^3_{1A}} \hat{\mathbf{l}}_{1A}, \ \ \hat{\mathbf{h}}_{\rm ee} = - \frac{\alpha^2}{2} \frac{1}{\hat{r}^3_{12}} \hat{\mathbf{l}}_{12},
\end{align}
and $\hat{\mathbf{T}}_{ij}$ ($\hat{T}^{x,y,z}_{ij}$) are the Cartesian triplet operators,
\begin{align}
    \hat{T}^x_{ij} & = \frac{1}{2} (\hat{a}^\dagger_{i\alpha} \hat{a}_{j\beta} + \hat{a}^\dagger_{i\beta} \hat{a}_{j\alpha}), \\
    \hat{T}^y_{ij} & = \frac{1}{2i} (\hat{a}^\dagger_{i\alpha} \hat{a}_{j\beta} - \hat{a}^\dagger_{i\beta} \hat{a}_{j\alpha}), \\
    \hat{T}^z_{ij} & = \frac{1}{2} (\hat{a}^\dagger_{i\alpha} \hat{a}_{j\alpha} - \hat{a}^\dagger_{i\beta} \hat{a}_{j\beta}).
\end{align}
Further discussion of the Breit-Pauli mean-field Hamiltonian can be found in Ref.~\cite{neese2005efficient}.

\subsection{XAS/RIXS spectra from correction vectors}

The XAS spectral function ($S$) and RIXS cross-section ($\sigma$) (averaged over all orientations, directions, and polarizations of the scattered radiation) can be written as
\begin{align}
S(\omega_{ex}) & = - \frac{1}{\pi} \Im \sum_{\lambda=x,y,z} \langle \Psi_0 | \hat{\mu}^{\dagger}_\lambda \frac{1}{\omega_{ex}-\hat{H}+E_0+i\eta} \hat{\mu}_\lambda | \Psi_0 \rangle, \label{eq:xas} \\
\sigma (\omega_{ex}, \omega_{em}) & = - \frac{16 \pi \omega_{em}^3 \omega_{ex}} {9 c^4} \sum_{\rho, \lambda = x, y, z} |\alpha_{\rho\lambda}(\omega_{ex}, \omega_{em})|^2,
\end{align}
with
\begin{align}
    |\alpha_{\rho \lambda}(\omega_{ex}, \omega_{em})|^2 
    = \Im & \left[ \langle \Psi_0 | \hat{\mu}^{\dagger}_\lambda \frac{1}{\omega_{ex}-\hat{H}+E_0-i\eta} \hat{\mu}_\rho \right. \nonumber \\
    & \left. \times \frac{1}{\omega_{ex}-\omega_{em}-\hat{H}+E_0+i\eta'} \hat{\mu}^{\dagger}_\rho 
            \frac{1}{\omega_{ex}-\hat{H}+E_0+i\eta} \hat{\mu}_\lambda | \Psi_0 \rangle \right] \label{eq:rixs}
\end{align}
and where $|\Psi_0\rangle$ and $E_0$ are the ground-state wavefunction and energy, $\hat{\boldsymbol{\mu}}$ is the dipole operator, $\omega_{ex}$ and $\omega_{em}$ are the energies of the incident and scattered radiation, and $\eta$ and $\eta'$ are Lorentzian broadening factors. We note that we only consider resonant terms in this work. 

The above expressions involve resolvent operators $\hat{R}(z) = [z- \hat{H}]^{-1}$. The correction vector approach to nonlinear properties~\cite{soos1989valence} involves computing the application of the resolvent to state $|C\rangle = \hat{R}|X\rangle$ by solving $(z-\hat{H})|C\rangle = |X\rangle$; $|C\rangle$ is termed the correction vector (CV). 

Following this, the XAS/RIXS quantities can  be computed in three steps: (1) solve for $|\Psi_0\rangle$, (2) solve for the response (from the correction vector equations), and (3) compute $S$/$\sigma$ from the correction vectors. Specifically, after obtaining 
$|\Psi_0\rangle$, we compute
the CVs ($\{ A_\lambda (\omega_{ex}) \}$) by solving \begin{align}
(\omega_{ex} - \hat{H} + E_0 + i\eta) |A_\lambda(\omega_{ex}) \rangle = 
\hat{\mu}_\lambda | \Psi_0 \rangle. \label{eq:rsp1}
\end{align}
and obtain the XAS spectral function
\begin{align}
    S(\omega_{ex}) & = - \frac{1}{\pi} \Im \sum_{\lambda=x,y,z} \langle \Psi_0 | \hat{\mu}^{\dagger}_\lambda 
| A_\lambda (\omega_{ex}) \rangle. \label{eq:xas_cv}
\end{align}
For the RIXS cross section, we solve for an additional set of CVs ($\{ B_{\rho\lambda} (\omega_{ex},\omega_{em}) \}$),
\begin{align}
(\omega_{ex} - \omega_{em} - \hat{H} + E_0 + i\eta') |B_{\rho\lambda} (\omega_{ex},\omega_{em}) \rangle 
& = \hat{\mu}^{\dagger}_\rho | A_\lambda (\omega_{ex}) \rangle, \label{eq:rsp2}
\end{align}
and compute the cross section using
\begin{align}
\sigma (\omega_{ex}, \omega_{em}) = - \frac{16 \pi \omega_{em}^3 \omega_{ex}} {9 c^4} \sum_{\rho, \lambda = x, y, z} \Im \langle A_\lambda (\omega_{ex}) | \hat{\mu}_\rho | B_{\rho\lambda} (\omega_{ex}, \omega_{em}) \rangle. \label{eq:rixs_cv}
\end{align}

\subsection{Interpretation of correction vectors}

The CVs can be formally expanded in a sum-over-states: 
\begin{align}
    |A_\lambda (\omega_{ex})\rangle & = \sum_I \frac{\langle \Psi_I | \hat{\mu}_\lambda | \Psi_0 \rangle} {\omega_{ex} - (E_I - E_0) + i\eta } |\Psi_I\rangle = \sum_I a_I |\Psi_I\rangle, \label{eq:a} \\
    |B_{\rho\lambda} (\omega_{ex},\omega_{em}) \rangle & =
    \sum_I \frac{\langle \Psi_I | \hat{\mu}^{\dagger}_\rho | A_\lambda (\omega_{ex}) \rangle}{\omega_{ex} - \omega_{em} - (E_I - E_0) + i\eta'} |\Psi_I\rangle = \sum_I b_I |\Psi_I\rangle, \label{eq:b}
\end{align}
where $\Psi_I$, $E_I$ are eigenstates, eigenvalues of $H$ respectively. 
On resonance  (i.e. a divergent denominator in Eq.~\eqref{eq:a}, Eq.~\eqref{eq:b}), $|\Psi_I\rangle$ is the final state. This has core-excited character in XAS and valence-excited character in RIXS. Note that final states in XAS are the intermediate states in RIXS.



\subsection{Deconvolution}

To interpret the L-edge XAS and RIXS spectra, we deconvolve the intermediates into particle-hole and spin contributions.

We define the particle-hole components for XAS ($S^{\rm (ph)}_{ia}(\omega_{ex})$) and for RIXS ($\sigma^{\rm (ph)}_{ia} (\omega_{ex}, \omega_{em})$) as
\begin{align}
    S^{\rm (ph)}_{ia}(\omega_{ex}) 
        & = - \frac{1}{\pi} \Im \sum_{\lambda=x,y,z} \mu_{\lambda, ia} \langle \Psi_0 |  \hat{a}^{\dagger}_i \hat{a}_a | A_\lambda (\omega_{ex}) \rangle, \label{eq:dcv_xas_ph} \\
    \sigma^{\rm (ph)}_{ia} (\omega_{ex}, \omega_{em})
        & = - \frac{16 \pi \omega_{em}^3 \omega_{ex}} {9 c^4} \sum_{\rho, \lambda = x, y, z} \sum_{jb} \mu_{\rho, ib} \mu_{\lambda, ja} \Im \langle A^{\rm (ph)}_{ja} (\omega_{ex}) | \hat{a}^{\dagger}_i \hat{a}_b | B_{\rho\lambda} (\omega_{ex}, \omega_{em}) \rangle, \label{eq:dcv_rixs_ph}
\end{align}
where the correction vector $A^{\rm (ph)}_{ia} (\omega_{ex})$ is obtained from
\begin{align}
    (\omega_{ex} - \hat{H} + E_0 + i\eta) |A^{\rm (ph)}_{ia} (\omega_{ex}) \rangle 
    = \hat{a}^{\dagger}_a \hat{a}_i | \Psi_0 \rangle.
\end{align}
On resonance, we interpret $S^{\rm (ph)}_{ia}$ as the amplitude of the core ($i$) to valence ($a$) excitation in the XAS final state, and $\sigma^{\rm (ph)}_{ia}$ as the amplitude of the valence ($i$) to valence ($a$) excitation in the RIXS final state.
In XAS, we further define the total valence (particle) contribution by summing over the core (hole) contributions
\begin{align}
	S^{\rm (p)}_{a}(\omega_{ex}) = \sum_i S^{\rm (ph)}_{ia}(\omega_{ex}). \label{eq:dcv_xas_p}
\end{align}

To deconvolve the spectra into  different spin contributions for XAS ($S^{\rm (s)}_S (\omega_{ex})$) and for RIXS ($\sigma^{\rm (s)}_S (\omega_{ex}, \omega_{em})$),
we apply spin projection operators ($P_S$),
\begin{align}
    S^{\rm (s)}_S (\omega_{ex}) & = - \frac{1}{\pi} \Im \sum_{\lambda=x,y,z} \langle \Psi_0 | \hat{\mu}^{\dagger}_\lambda P_S | A_\lambda (\omega_{ex}) \rangle, \label{eq:dcv_xas_s2} \\
    \sigma^{\rm (s)}_S (\omega_{ex}, \omega_{em}) 
    & = - \frac{16 \pi \omega_{em}^3 \omega_{ex}} {9 c^4} \sum_{\rho, \lambda = x, y, z} \Im \langle A_\lambda (\omega_{ex}) | \hat{\mu}_\rho P_S | B_{\rho\lambda} (\omega_{ex}, \omega_{em}) \rangle. \label{eq:dcv_rixs_s2}
\end{align}
We use here L\"{o}wdin's spin projector,~\cite{lowdin1955quantum} 
\begin{align}
    P_S = \prod_{S'\ne S} \frac{\hat{S}^2 - S'(S'+1)}{S(S+1) - S'(S'+1)}.
\end{align}

For all deconvolution schemes, the sum of the deconvolved spectra is the total spectrum.
\begin{align}
    S (\omega_{ex}) 
      & = \sum_{ia} S_{ia}^{\rm (ph)} (\omega_{ex}) 
        = \sum_{a} S_a^{\rm (p)} (\omega_{ex}) 
        = \sum_{S} S_S^{\rm (s)} (\omega_{ex}) \\
    \sigma (\omega_{ex}, \omega_{em}) 
      & = \sum_{ja} \sigma^{\rm (ph)}_{ja} (\omega_{ex}, \omega_{em})
         = \sum_{S} \sigma^{\rm (s)}_{S} (\omega_{ex}, \omega_{em})
\end{align}

\section{Computational Details} \label{sec:comput}

We take the geometries of $\ce{[FeCl_4]^{1-/2-}}$ and $\ce{[Fe(SCH_3)_4]^{1-/2-}}$ from previous computational studies~\cite{maganas2019comparison, chilkuri2017revisiting}.
The $\ce{[Fe(SCH_3)_4]^{1-}}$ geometry comes from the X-ray crystal structure, and the geometries of other complexes correspond to optimized DFT structures. 
We used the restricted active space (RAS) ansatz for the ground-state wavefunction ($\Psi_0$) and the correction vectors ($A_\lambda$ and $B_{\lambda, \rho}$) (see Sec.~\ref{sec:act}).
 The RAS ansatz is implemented using matrix product state (MPS) techniques (see Sec.~\ref{sec:mps}). 
We construct the active space using a procedure described in Sec.~\ref{sec:orb}. 
For the spectra calculations, we used Lorentzian broadening factors of $\eta=0.3$ eV and $\eta'=0.1$ eV  in Eqs. (\ref{eq:xas}) and (\ref{eq:rixs}), respectively.
We simulated RIXS spectra using the incident radiation energy ($\omega_{ex}$) that results in the maximum intensity of the L${}_3$ band in the XAS spectra.
All calculations were performed using the PySCF~\cite{sun2015libcint, sun2018pyscf, sun2020recent} and Block2~\cite{block2} packages.

\subsection{Active space models} \label{sec:act}

A RAS ansatz can be written in occupation number form as
\begin{align}
    |A \rangle = \sum_{m_1 m_2 ... m_{M}} \sum_{n_1 n_2 ... n_{N}} a_{m_1 m_2 ... m_M n_1 n_2 ... n_N} | 1 1 ... 1 m_1 m_2 ... m_M n_1 n_2 ... n_N 0 0 ... 0\rangle \label{eq:ras}
\end{align}
where $m_i$ and $n_j$ are the occupations ($0,1$) in the two active subspaces, RAS1 and RAS2, respectively, and $a_{m_1 m_2 ... m_M n_1 n_2 ... n_N}$ is the coefficient of the determinant $| 1 1 ... 1 m_1 m_2 ... m_M n_1 n_2 ... n_N 0 0 ... 0\rangle$.
RAS1 consists of $M$ occupied orbitals with a maximum number of holes ($M_{\rm hole}$), i.e. $\sum_{i=1}^M m_i \geq M - M_{\rm hole}$. 
RAS2 consists of $N$ orbitals with no restrictions on the electron occupancy, except for the total number of electrons in the RAS spaces, i.e., $\sum_{i=1}^M m_i + \sum_{i=1}^N n_i = N^{\rm RAS}_{\rm elec}$.
Additional RAS partitions can be introduced.

For the XAS spectra, we used minimal RAS models, with five 3$d$ valence orbitals of $\ce{Fe}$ in RAS2 and three 2$p$ core orbitals of $\ce{Fe}$ in RAS1, with $M_{\rm hole}=1$, which we designate as RAS1($2p_{\mathrm{Fe}}^6$)RAS2($3d_{\mathrm{Fe}}^{5/6}$).
For the RIXS spectra, we also considered larger active space models with an additional RAS1 (RAS1') partition, consisting of 
four $\sigma$-bonding orbitals of the $\ce{Fe-Cl}$ or $\ce{Fe-S}$ bonds; both RAS1/RAS1' have $M_{\rm hole}=1$.
We designate this ansatz RAS1($2p_{\mathrm{Fe}}^6)$RAS1'($ \sigma_{\ce{Fe-Cl/S}}^8$)RAS2($3d_{\mathrm{Fe}}^{5/6}$).

\subsection{Matrix product state (MPS) implementation} \label{sec:mps}

We implement the RAS ansatz within the matrix product state formalism.  We rewrite the  configuration coefficients in Eq.~\ref{eq:ras} as
\begin{align}
    a_{m_1 m_2 ... m_M n_1 n_2 ... n_N} 
    = \sum_{d_1, d_2, ..., d_{M+N}} 
    A^{m_1}_{d_1} A^{m_2}_{d_1 d_2} ... A^{m_M}_{d_{M-1} d_{M}}
    A^{n_1}_{d_{M}d_{M+1}} A^{n_2}_{d_{M+1}d_{M+2}} ... A^{n_N}_{d_{M+N-1}} \label{eq:mps}
\end{align}
where $d_i$ is a bond index of dimension ($D$), 
$\mathbf{A}^{m_k}$ ($1<k\le M$) and $\mathbf{A}^{n_k}$ ($1\le k<N$) are $D\times D$ matrices, and $\mathbf{A}^{m_1}$ and $\mathbf{A}^{n_N}$ are $1\times D$ and $D\times 1$ vectors, respectively. 
All the matrices (and vectors) contain complex elements. Here we choose the bond dimensions $D$ so that the RAS ansatz is exactly represented: there is no MPS compression, and the MPS formalism is only used to simplify the implementation.

To compute ground-states we used the density matrix renormalization group (DMRG) algorithm~\cite{chan2002highly, chan2004algorithm, chan2009density}, and we used the dynamical DMRG algorithm~\cite{ronca2017time} to solve the correction vector equations in Eqs.~\ref{eq:rsp1} and~\ref{eq:rsp2}.


\subsection{Active space construction} \label{sec:orb}

To construct the active space for the RAS ansatz,
we used a technique introduced in a previous study of iron-sulfur clusters.~\cite{li2019electronic} 
We first performed unrestricted DFT calculations for the high-spin state without spin-orbit coupling using the BP86 functional \cite{becke1988density, perdew1986density} in the ANO-RCC-VDZP basis~\cite{roos2005new}.
Then, we computed unrestricted natural orbitals as the eigenvectors of the sum of alpha and beta DFT density matrices. We then identified active space orbitals from the unrestricted natural orbital occupation numbers, and localized the orbitals within the active space to improve the convergence of DMRG and dynamic DMRG algorithms.

\section{Results and Discussion} \label{sec:results}

Using the above formalism, we simulated the L-edge XAS and 2p3d RIXS spectra of the $\ce{[Fe^{II/III}Cl_4]^{2-/1-}}$ and $\ce{[Fe^{II/III}(SCH_3)_4]^{2-/1-}}$ complexes. 
We 
compare our results to experimental spectra, normalizing the maximum intensity of the L${}_3$-edge band of XAS to $1$ and that of the highest intensity band of RIXS to 0.2 (for the ferrous complexes), 0.8 (for the tetrachloride ferric complex), and 0.6 (for the tetrathiolate ferric complex).
The experimental data are taken from Refs.~\citenum{kowalska2017iron, maganas2019comparison} for XAS and Ref.~\citenum{van2018electronic} for RIXS.
Note that for the tetrathiolate ferrous and ferric complexes, we used experimental spectra of complexes with benzenethiolate (SPh) and 2,3,5,6-tetramethylbenzenethiolate (SDur) ligands, rather than the methylthiolate ligands used in our computations.

\subsection{L-edge XAS of $\ce{[Fe^{II}Cl_4]^{2-}}$ and $\ce{[Fe^{II}(SCH_3)_4]^{2-}}$} \label{sec:xas_feII}

\begin{figure}[ht!]
\center{\includegraphics[width=0.8\columnwidth]{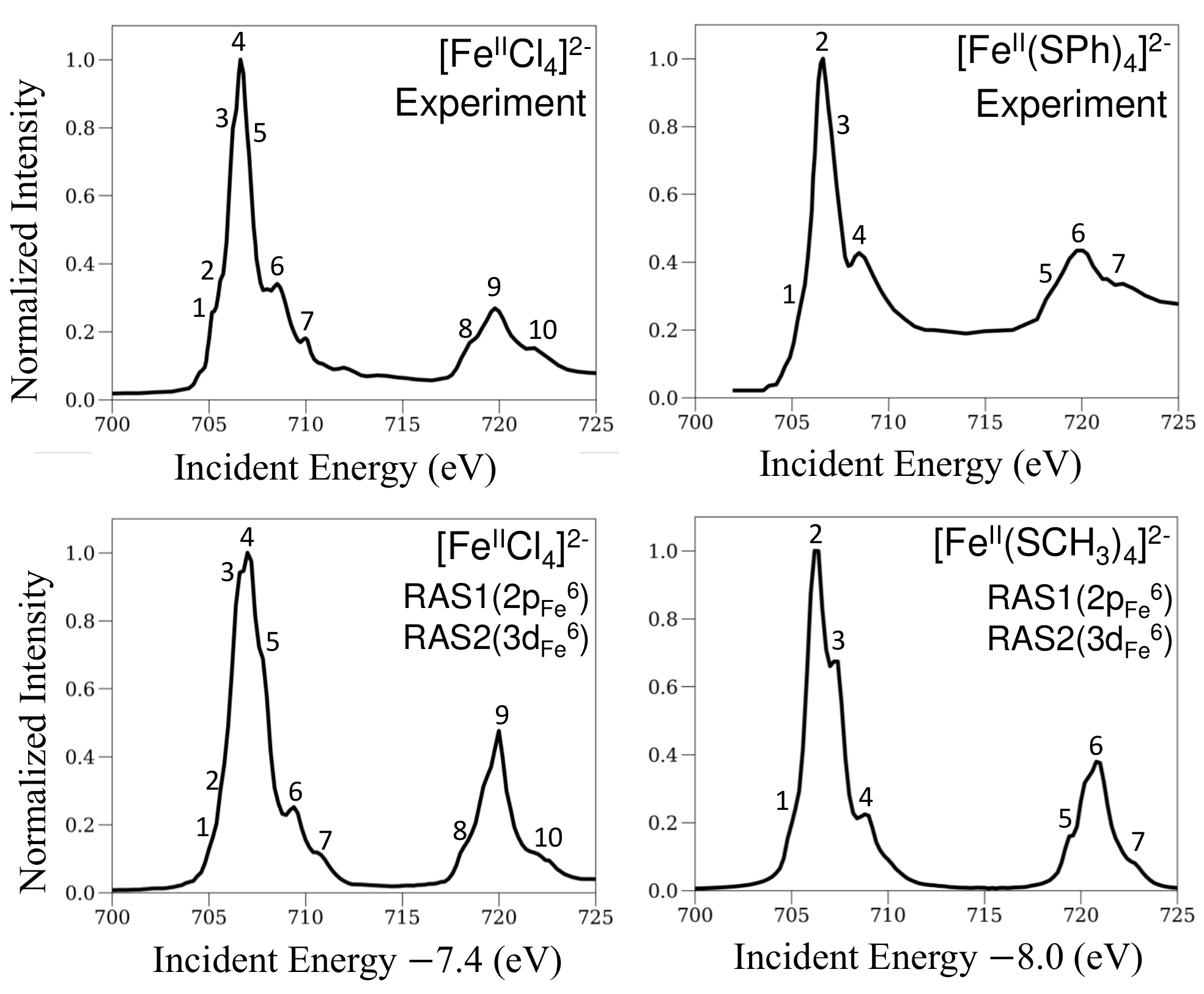}}
\caption{L-edge XAS spectra of ferrous tetrachloride and tetrathiolate complexes, left and right panels, respectively. The experimental spectra in the upper panels are taken from Refs.~\citenum{maganas2019comparison, kowalska2017iron}. Important features of the experimental spectra are enumerated based on earlier experimental studies~\cite{maganas2019comparison, kowalska2017iron} and the corresponding features in the theoretical spectra are enumerated.
}\label{fig:xas_feII}
\end{figure}

We first discuss the simulations of the L-edge XAS spectra for the ferrous complexes $\ce{[Fe^{II}Cl_4]^{2-}}$ and $\ce{[Fe^{II}(SCH_3)_4]^{2-}}$ using  the RAS1($2p_{\mathrm{Fe}}^6$)RAS2($3d_{\mathrm{Fe}}^{6/5}$) active space model (see Sec.~\ref{sec:act}).
Fig.~\ref{fig:xas_feII}, shows the experimental (upper panels) and calculated (lower panels) spectra of $\ce{[Fe^{II}Cl_4]^{2-}}$ (left panels) and $\ce{[Fe^{II}(SCH_3)_4]^{2-}}$ (right panels).
The theoretical spectra with the minimal RAS model are in good agreement with the experimental spectra, capturing the relative intensities and relative energy positions of the bands.
However, the theoretical spectra for $\ce{[Fe^{II}Cl_4]^{2-}}$ and $\ce{[Fe^{II}(SCH_3)_4]^{2-}}$ are shifted by $7.4$ and $8.0$ eV, respectively, ($\sim 10\%$ error) compared to the experimental spectra.
Previous studies using the MRCI method with a single excitation amplitude~\cite{maganas2019comparison} also reported a constant energy-shift error of size $8$, $3$, and $2.7$ eV as the size of the cc-pwCVXZ basis set was increased from X = D, T, to Q, while other studies have shown that orbital relaxation also reduces the constant energy shift error.~\cite{hait2021orbital}
Thus we attribute our constant shift error to the small size and lack of orbital relaxation in the RAS model. 

\begin{figure}[ht!]
\center{\includegraphics[width=0.9\columnwidth]{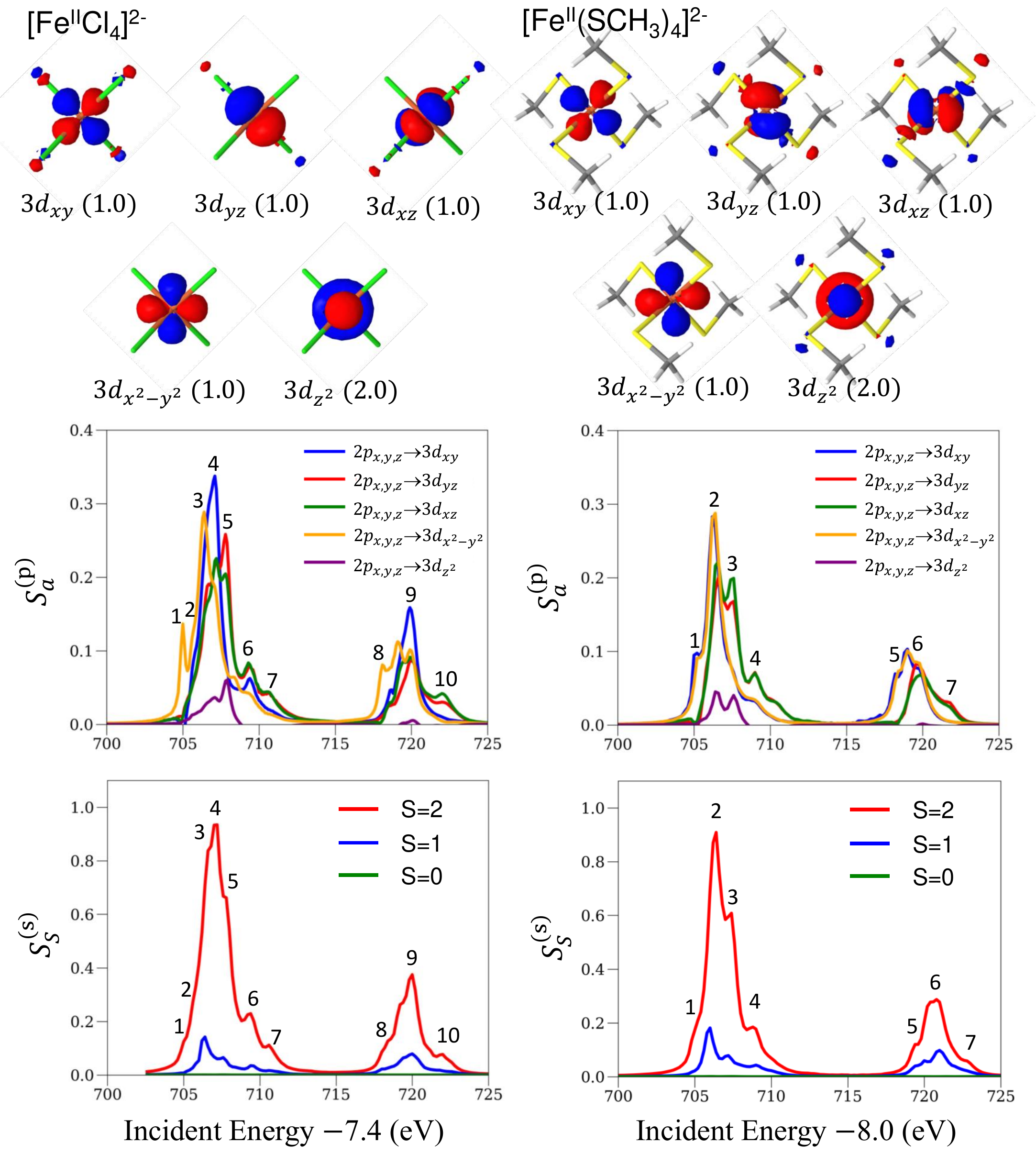}}
\caption{Natural orbitals of the ground state RAS wave function (top panels) and deconvolved XAS spectra for particle (middle panels) and spin (bottom panels) contributions. $\ce{[Fe^{II}Cl_4]^{2-}}$ (left panel) and $\ce{[Fe^{II}(SCH_3)_4]^{2-}}$ (right panel). The particle contributions are measured in the natural orbital basis shown in the top panels. }\label{fig:xas_feII_deconv}
\end{figure}

We show deconvolved spectra for $\ce{[Fe^{II}Cl_4]^{2-}}$ and $\ce{[Fe^{II}(SCH_3)_4]^{2-}}$ in the left and right panels in Fig.~\ref{fig:xas_feII_deconv}, respectively.
The particle-hole deconvolution was done in the natural orbital basis 
 of the ground-state RAS wavefunction,
and the valence natural orbitals (labelled by atomic orbital character) are shown in the top panels of Fig.~\ref{fig:xas_feII_deconv}. 
The values in parentheses for each natural orbital denotes the natural occupation number.
In the middle panels of Fig.~\ref{fig:xas_feII_deconv}, we present the particle (valence) contribution, as defined by summing over the core indices in the the core-to-valence decomposition in Eq.~\ref{eq:dcv_xas_p}.
Each representative band is observed to result from a different particle (valence) contribution,
and based on their band positions,  
we can conclude that the approximate orbital energy order is $3d_{z^2} < 3d_{x^2-y^2} < 3d_{xy} < 3d_{yz} = 3d_{xz}$ for $\ce{[Fe^{II}Cl_4]^{2-}}$ and $3d_{z^2} < 3d_{x^2-y^2} = 3d_{xy} < 3d_{yz} = 3d_{xz}$ for $\ce{[Fe^{II}(SCH_3)_4]^{2-}}$. 

In the bottom panels of Fig.~\ref{fig:xas_feII_deconv}, we present the deconvolved spectra for the different spin components. 
For both \ce{[Fe^{III}Cl_4]^{1-}} and \ce{[Fe^{III}(SCH_3)_4]^{1-}}, the largest contribution is for the same spin-component as the (high-spin) ground state ($S=2$), with contributions of $88 \%$ and $83 \%$, respectively. The contributions of the $\Delta S = 1$ transitions are $12 \%$ and $17 \%$, while those of the $\Delta S = 2$ transitions are negligible.
Interestingly, the percentage of the $\Delta S = 1$ contribution is similar for both complexes, possibly due to similar spin-orbit coupling strengths resulting from the same oxidation state.

\subsection{L-edge XAS of \ce{[Fe^{III}Cl_4]^{1-}} and \ce{[Fe^{III}(SCH_3)_4]^{1-}}} \label{sec:xas_feIII}

\begin{figure}[ht!]
\center{\includegraphics[width=0.8\columnwidth]{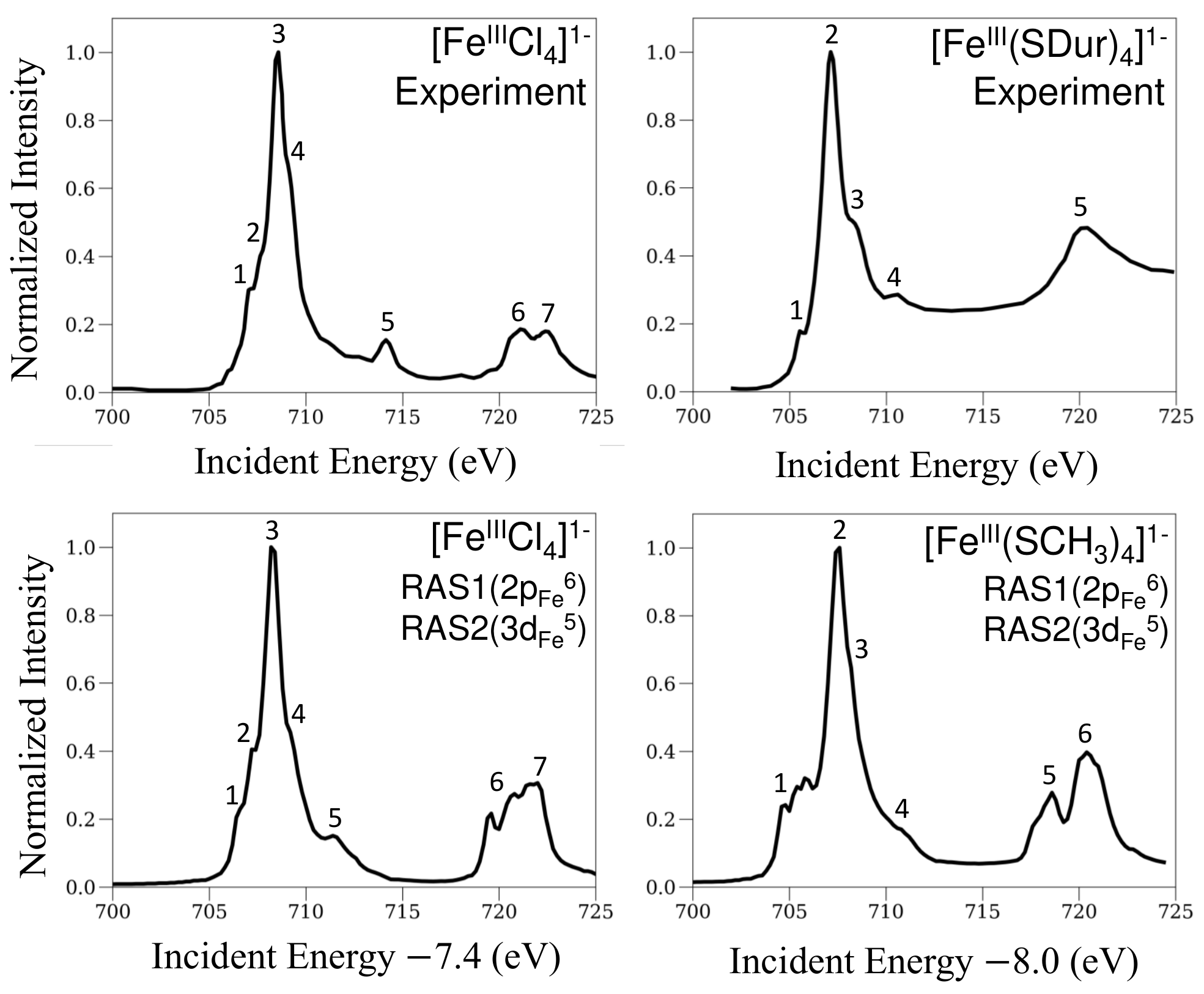}}
\caption{L-edge XAS spectra of ferric tetrachloride and tetrathiolate complexes, left and right panels, respectively, with the same format as Fig.~\ref{fig:xas_feII}.  }\label{fig:xas_feIII}
\end{figure}

\begin{figure}[ht!]
\center{\includegraphics[width=0.9\columnwidth]{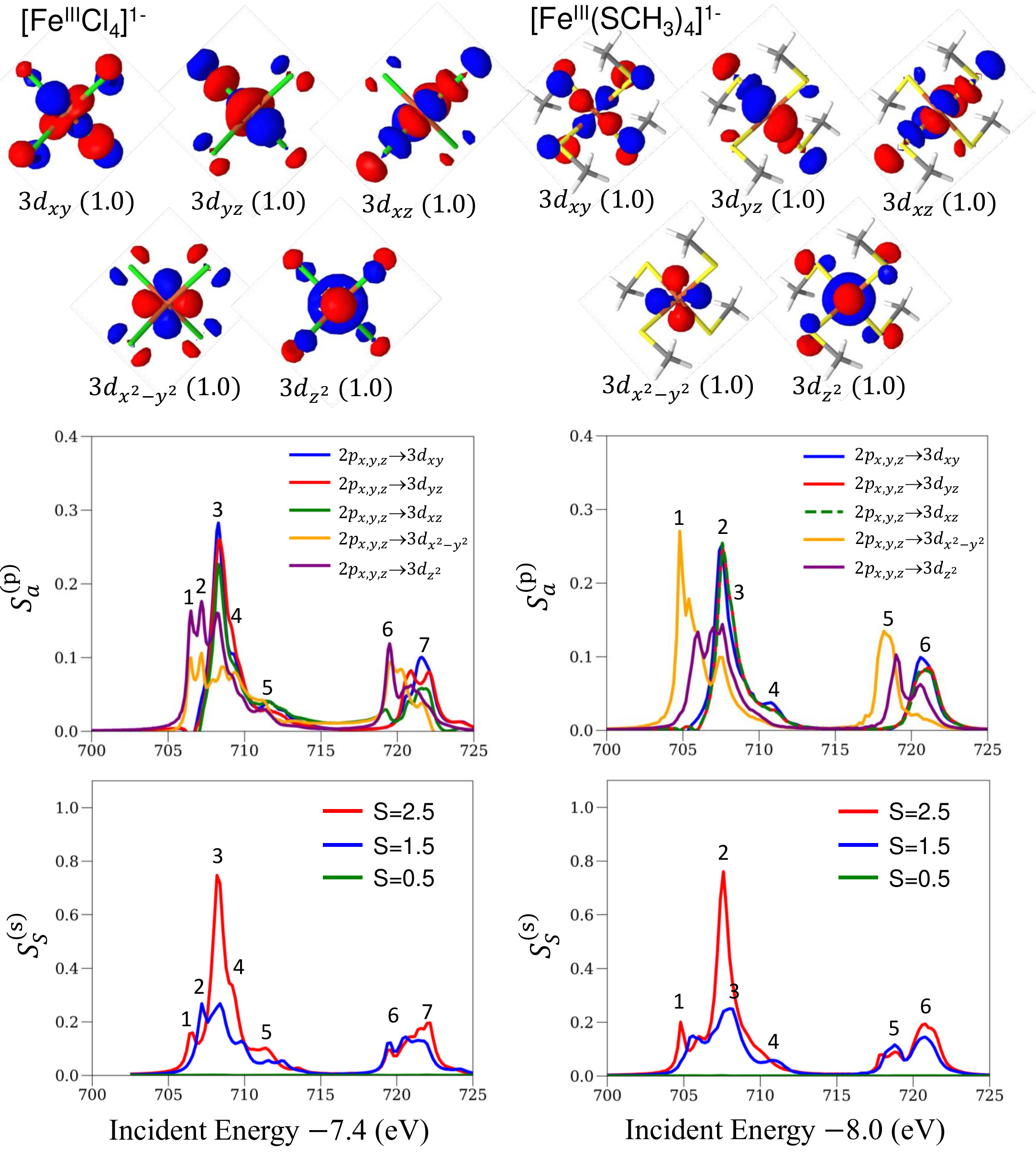}}
\caption{Natural orbitals of the ground state RAS wave function (top panels) and deconvolved XAS spectra for particle (middle panels) and spin contributions (bottom panels). $\ce{[Fe^{III}Cl_4]^{1-}}$ (left panel) and $\ce{[Fe^{III}(SCH_3)_4]^{1-}}$ (right panel). The particle contributions are measured in the natural orbital basis shown in the top panels.
}\label{fig:xas_feIII_deconv}
\end{figure}

We next present XAS spectra of the ferric complexes in Fig.~\ref{fig:xas_feIII}.
The theoretical spectra of \ce{[Fe^{III}Cl_4]^{1-}} and \ce{[Fe^{III}(SCH_3)_4]^{1-}} have similar constant energy-shift errors of $7.4$ and $8.0$ eV, respectively.
The general features of the theoretical XAS spectra are in good agreement with the experimental ones.
However,  the relative band positions do not match up as well as in the ferrous complexes.
For example, in \ce{[Fe^{III}Cl_4]^{1-}}, the energy position of the fifth band is shifted relative to the highest peak by $2.5$ eV. 
In an earlier MRCI and MREOM-CC study~\cite{maganas2019comparison}, this shift has been attributed to multi-reference electron correlation effects involving the ligand $3p$ orbitals and metal $3d$ and $4s$ orbitals.
In \ce{[Fe^{III}(SCH_3)_4]^{1-}}, in the theoretical spectrum, the first band has a broader shoulder, the third band is closer to the second band, and the L${}_2$ edge
has two clear bands, as opposed to only one in the experiment. 
These differences can be better understood using the deconvolved spectra in Fig.~\ref{fig:xas_feIII_deconv}, which we now discuss.

The top panels of Fig.~\ref{fig:xas_feIII_deconv} represent the valence orbitals of the natural orbitals for the ground-state RAS wave functions. 
In contrast to the ferrous complexes, the natural orbitals of the ferric complexes have significant mixing with $3p$ orbitals of the ligands, due to the 
 higher oxidation state of Fe.
An exception to this is the $3d_{x^2-y^2}$ orbital of $\ce{[Fe^{III}(SCH_3)_4]^{1-}}$, which has little mixing.
The middle panels of Fig.~\ref{fig:xas_feIII_deconv} shows the particle (valence) contributions. 
In $\ce{[Fe^{III}Cl_4]^{1-}}$ and $\ce{[Fe^{III}(SCH_3)_4]^{1-}}$, the band positions of $3d_{xy}$, $3d_{yz}$, and $3d_{xz}$ are identical.
In \ce{[Fe^{III}Cl_4]^{1-}}, the band positions of $3d_{x^2-y^2}$ and $3d_{z^2}$ are also the same.
Thus, the approximate orbital energy order is $3d_{z^2} = 3d_{x^2-y^2} < 3d_{xy} = 3d_{yz} = 3d_{xz}$, in agreement with an ideal tetrahedral ligand.
However, in \ce{[Fe^{III}(SCH_3)_4]^{1-}}, the band positions of $3d_{x^2-y^2}$ are shifted by $-2$ eV relative to those of $3d_{z^2}$. The shift in the $3d_{x^2-y^2}$ bands is the main reason for the disagreement between the theoretical and experimental spectra in the right panels of Fig.~\ref{fig:xas_feIII}.
The small mixing between the $3d_{x^2-y^2}$ orbital and the $3p$ ligand orbitals in Kohn-Sham density functional theory, used to construct the active space, contributes to this shift. 

In the bottom panels, we present the deconvolved spectra for the different spin components. 
For both complexes, the largest contribution ($60 \%$) comes from states with the same spin as the (high-spin) ground states ($S=2.5$).
Contributions of $40 \%$ are observed for the $\Delta S = 1$ transition for both ferric complexes, while the contribution of the $\Delta S = 2$ transition is negligible.
The higher contributions for the $\Delta S = 1$ transition in the ferric complexes, as compared to the ferrous complexes, can be attributed to the stronger spin-orbit coupling due to the higher oxidation state of Fe.

\subsection{RIXS spectra of $\ce{[Fe^{II}Cl_4]^{2-}}$ and $\ce{[Fe^{II}(SCH_3)_4]^{2-}}$}

\begin{figure}[ht!]
\center{\includegraphics[width=\columnwidth]{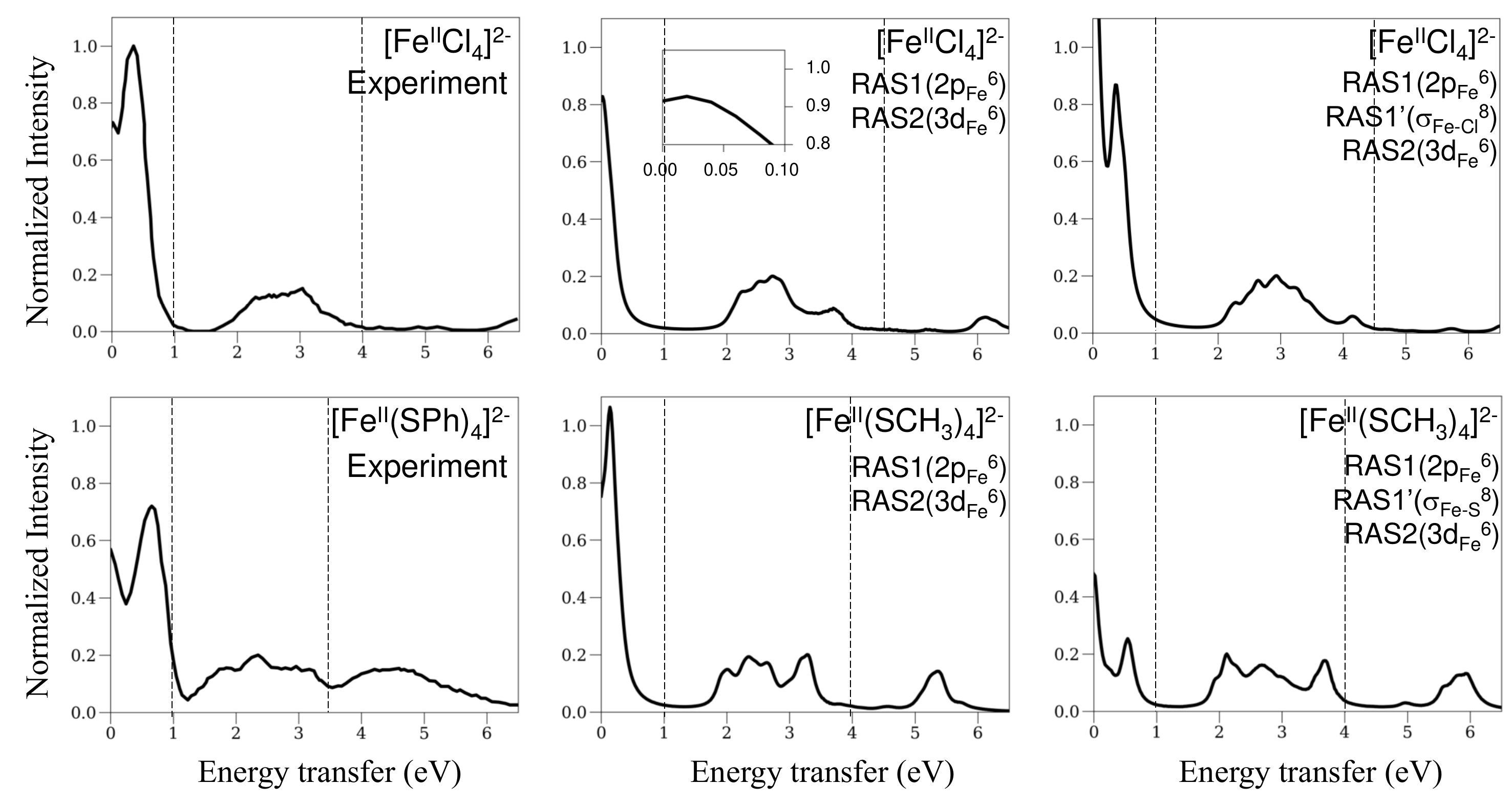}}
\caption{2p3d RIXS spectra of $\ce{[Fe^{II}Cl_4]^{2-}}$ and $\ce{[Fe^{II}(SCH_3)_4]^{2-}}$ in upper and lower panels, respectively. The experimental spectra in the left panels are from Ref.~\citenum{van2018electronic}.  Theoretical spectra with two different active space models are in shown in the center and right panels, respectively. }\label{fig:rixs_feII}
\end{figure}

Next, we discuss the 2p3d RIXS spectra of the ferrous complexes, $\ce{[Fe^{II}Cl_4]^{2-}}$ and $\ce{[Fe^{II}(SCH_3)_4]^{2-}}$. 
Figure~\ref{fig:rixs_feII} shows the experimental spectra (left panels) and theoretical spectra (center and right panels) with different active space models (see Sec.~\ref{sec:act}).
For convenience, we divide the spectra into three regions divided by the black dashed lines.
In the first region (0--1 eV), there are two bands at 0.00 and 0.35 eV for $\ce{[Fe^{II}Cl_4]^{2-}}$ and at 0.00 and 0.67 eV for $\ce{[Fe^{II}(SCH_3)_4]^{2-}}$ in the experimental spectra.
However, we observed only one band in the theoretical spectra using the minimal active space, at 0.02 eV for $\ce{[Fe^{II}Cl_4]^{2-}}$ and 0.14 eV for $\ce{[Fe^{II}(SCH_3)_4]^{2-}}$.
If we change to the larger 
RAS1($2p_{\mathrm{Fe}}^6)$RAS1'($ \sigma_{\ce{Fe-Cl/S}}^8$)RAS2($3d_{\mathrm{Fe}}^{5/6}$) active space (that
includes the the four occupied sigma-bonding orbitals between the Fe and Cl/S atoms) two bands correctly appear, at 0.00 and 0.36 eV for $\ce{[Fe^{II}Cl_4]^{2-}}$ and 0.00 and 0.54 eV for $\ce{[Fe^{II}(SCH_3)_4]^{2-}}$. 
This emphasizes the importance of electron correlation between the $3d$ orbitals and the sigma-bonding orbitals to reproduce the energy splitting in this energy range.

In the second energy region (1--4 eV), the experimental spectra contains broad bands, with half maximum intensity (HM) at 2.10 and 3.31 eV for the chloride complex, and 1.49 and 3.45 eV for the thiolate complex, and with full widths at half maximum (FWHM) of 1.21 and 1.96 eV, respectively.
The minimal active space model has narrower bands with HM at 2.12 and 3.07 eV (chloride complex) and 1.86 and 3.42 eV (thiolate complex), and FWHM of 0.95 and 1.56 eV.
Including the sigma-bonding orbitals broadens the bands with HM at 2.24 and 3.50 eV (chloride complex) and 1.88 and 3.84 eV (thiolate complex) and FWHM of 1.26 and 1.96 eV. 
These latter results are closer to what is seen in experiment, but are slightly shifted to positive energy.

In the third region (4--6 eV) of the experimental spectra, there is no representative band for the chloride complex, while there is a broad band in a range of 3.5--6 eV for the thiolate complex.
The corresponding band in the theoretical spectra for the thiolate complex is much narrower than the experimental band. We ascribe the difference to missing certain important states,  such as the ligand-to-metal charge transfer (LMCT) states, in the active space models.

\begin{figure}[ht!]
\center{\includegraphics[width=\columnwidth]{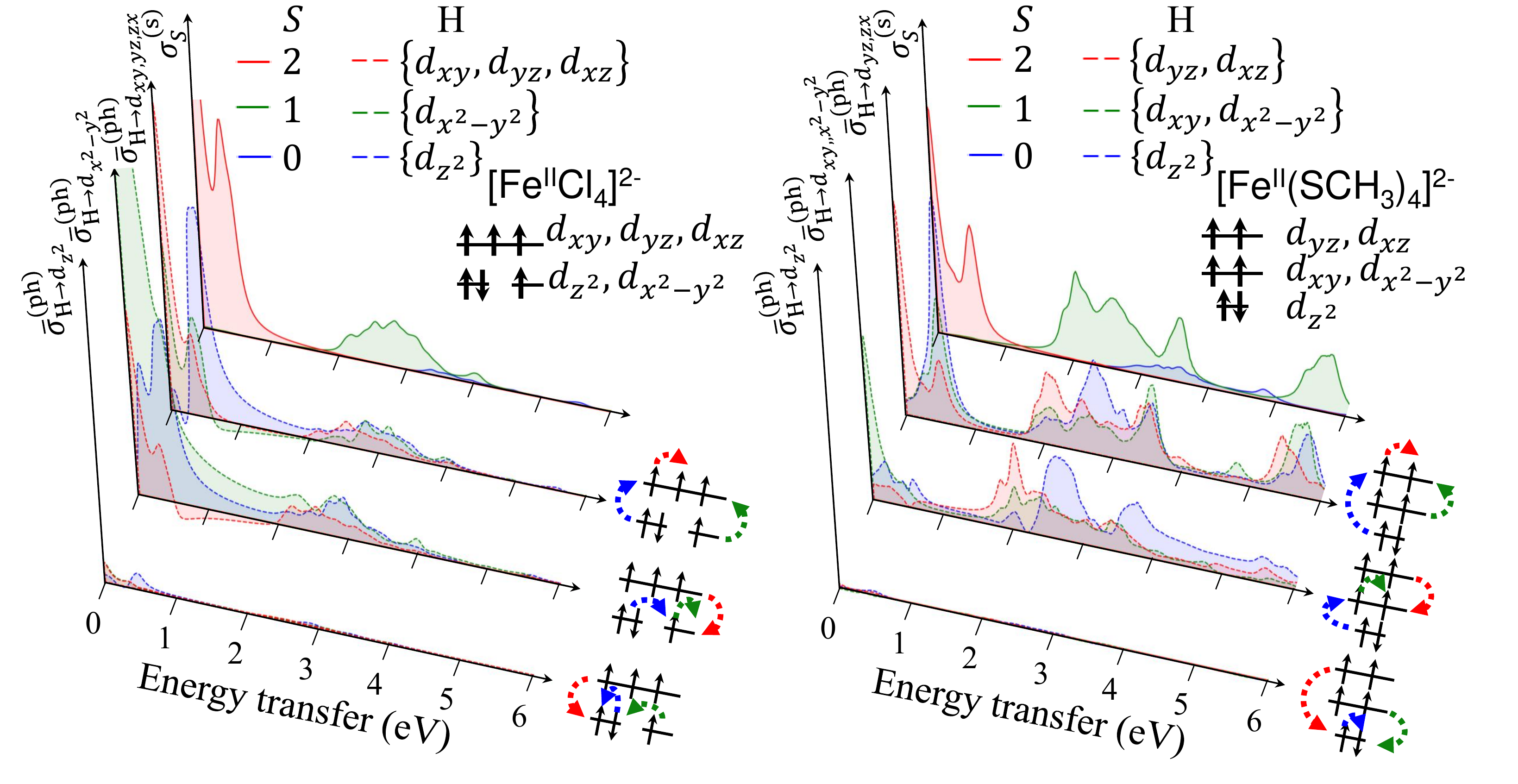}}
\caption{ Deconvolved theoretical RIXS spectra for \ce{[Fe^{II}Cl_4]^{2-}} (left panels) and \ce{[Fe^{II}(SCH_3)_4]^{2-}} complexes (right panels)  using the larger active space model.
Each panel shows four deconvolved spectra.
From front to back: 
The first three show the average particle-hole (valence-to-valence excitation) contributions ($\bar{\sigma}^{\rm (ph)}_{\mathrm{H}\rightarrow \mathrm{P}}$ in Eq.~\ref{eq:avg_ph_dcv}) for three particle sets ($\mathrm{P}$) (scale is in arbitrary units).
In each spectrum, the different contributions of the three hole sets ($\mathrm{H}$) are represented by dashed lines of different colors, and the corresponding transitions are depicted in the orbital diagram next to the x-axis by dashed arrows (using the same color scheme) next to the x-axis. 
The inset depicts the approximate ground state electronic configuration and the  three sets of $3d$ orbitals used to compute the average particle-hole contributions.
The remaining graph at the rear represents the spin-state contribution to the final states ($\sigma^{\rm (s)}_S$) defined in Eq.~\ref{eq:dcv_rixs_s2}. 
}\label{fig:feII_rixs_deconv}
\end{figure}

We further analyze the RIXS spectra of the ferrous complexes (in the larger active space model)
by deconvolving them, as shown in Fig.~\ref{fig:feII_rixs_deconv}.
In each panel, the spectrum in the rearmost position shows the spin-state deconvolution defined in Eq.~\ref{eq:dcv_rixs_s2}.
The ground state for both complexes is a spin quintet ($S=2$). Thus, the red curve, which depicts quintet contributions, shows the spin-allowed transitions, while the green and blue curves depict spin-forbidden transitions with $\Delta S = 1, 2$, respectively. 
The remaining three spectra from front to back represent the average particle-hole contributions defined as follows. (To simplify the analysis, we considered only valence-to-valence transitions, which account for over $99 \%$ of the total spectrum).
In addition, we partition the five $3d$ orbitals into three sets:  
$\{ d_{z^2} \}$, $\{ d_{x^2-y^2} \}$, $\{ d_{xy}, d_{yz}, d_{xz} \}$ for \ce{[Fe^{II}Cl_4]^{2-}} and $\{ d_{z^2} \}$, $\{ d_{x^2-y^2}, d_{xy} \}$, $\{ d_{yz}, d_{xz} \}$ for \ce{[Fe^{II}(SCH_3)_4]^{2-}}.
The average particle-hole contribution is defined by
\begin{align}
	\bar{\sigma}^{\rm (ph)}_{\mathrm{H}\rightarrow \mathrm{P}} = \frac{1}{N_\mathrm{H} N_\mathrm{P}} \sum_{i \in \mathrm{H}} \sum_{j \in \mathrm{P}} \sigma^{\rm (ph)}_{ij}, \label{eq:avg_ph_dcv}
\end{align}
where $N_\mathrm{H}$ and $N_\mathrm{P}$ are the number of orbitals in the $\mathrm{H}$ and $\mathrm{P}$ sets and $\sigma^{\rm (ph)}_{ij}$ is defined in Eq.~\ref{eq:dcv_rixs_ph}. For example, $N_{d_{z^2}}= 1$ and $N_{d_{xy, yz, zx}} = 3$ for \ce{[Fe^{II}Cl_4]^{2-}} giving $\bar{\sigma}^{\rm (ph)}_{d_{z^2} \rightarrow d_{xy, yz, zx}} = \frac{1}{3} (\sigma^{\rm (ph)}_{d_{z^2} d_{xy}}+\sigma^{\rm (ph)}_{d_{z^2} d_{yz}}+\sigma^{\rm (ph)}_{d_{z^2} d_{xz}})$.

The deconvolved spectra for the \ce{[Fe^{II}Cl_4]^{2-}} complex are shown in the left panel of Fig.~\ref{fig:feII_rixs_deconv}.
Throughout the energy range 0--6 eV, the main contributions come from transitions to $d_{x^2-y^2}$ and $d_{xy, yz, xz}$. 
The two dominant bands in the 0--1 eV range originate from spin-allowed transitions 
of $d_{z^2} \rightarrow d_{x^2-y^2}$ and $d_{z^2} \rightarrow d_{xy, yz, xz}$.
We also observe minor contributions from other transitions, which we attribute to the multi-reference nature of the quintet excited states, as they cannot be described by a single electronic configuration. 
The broad band in the energy range 1--4 eV primarily results from spin-forbidden transitions with $\Delta S = 1$. 
The spin-flip transitions of $d_{xy, yz, xz} \rightarrow d_{xy, yz, xz}, d_{x^2-y^2}$ 
and 
$d_{x^2-y^2} \rightarrow d_{x^2-y^2}$ 
contribute to the lower energy region of the broad band,
while the spin-flip transitions of $d_{x^2-y^2} \rightarrow d_{xy, yz, xz}, d_{x^2-y^2}$ 
and $d_{z^2} \rightarrow d_{xy, yz, xz}, d_{x^2-y^2}$ 
contribute to the higher energy region of the band.
The $\Delta S = 2$ contribution suggests the presence of double spin-flip excitations.

The right panel of Fig.~\ref{fig:feII_rixs_deconv} shows the deconvolved spectra for the \ce{[Fe^{II}(SCH_3)_4]^{2-}} complex.
As with the chloride complex, the main transitions are to $d_{x^2-y^2}$ and $d_{xy, yz, xz}$.
The bands in the 0--1 eV range are primarily due to spin-allowed transitions of $d_{z^2} \rightarrow d_{yz, xz}$ (maximum intensity at 0.54 eV) and minor transitions of $d_{z^2} \rightarrow d_{xy, x^2-y^2}$ (maximum intensity at 0.20 eV).
Like the chloride complex, spin-forbidden transitions of $\Delta S = 1$ mainly contribute to the broad band in the 1--4 eV range, with a minor contribution from $\Delta S = 2$ transitions in the 3--4 eV range.
The band can be further divided into three parts based on the dominant spin-flip transitions: transitions of $d_{yz, xz} \rightarrow d_{xy, x^2-y^2}, d_{yz, zx}$ and $d_{xy, x^2-y^2} \rightarrow d_{xy, x^2-y^2}, d_{yz, zx}$, transitions of $d_{z^2} \rightarrow d_{xy, x^2-y^2}, d_{yz, zx}$, and transitions of $d_{x^2-y^2, xy}, d_{yz, xz}, d_{z^2} \rightarrow d_{yz, xz}$ and $d_{z^2} \rightarrow d_{xy, x^2-y^2}$.
In the energy range of 4--6 eV, the $d-d$ contributions are similar to those in the higher part of  the broad band in the 3--4 eV range.

\subsection{RIXS spectra of $\ce{[Fe^{III}Cl_4]^{1-}}$ and $\ce{[Fe^{III}(SCH_3)_4]^{1-}}$}

\begin{figure}[ht!]
\center{\includegraphics[width=\columnwidth]{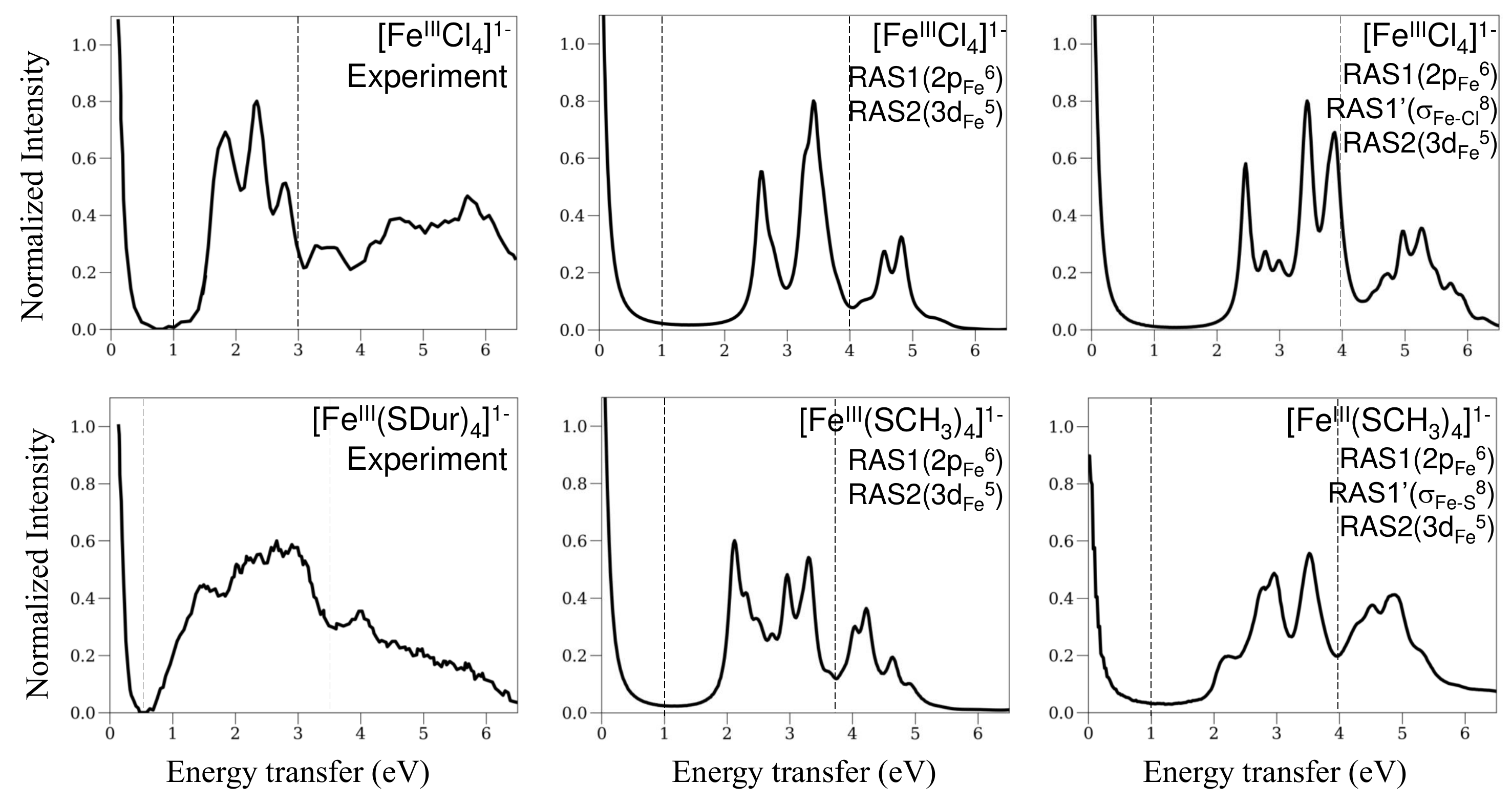}}
\caption{2p3d RIXS spectra of $\ce{[Fe^{III}Cl_4]^{1-}}$ and $\ce{[Fe^{III}(SCH_3)_4]^{1-}}$ in upper and lower panels, respectively, with the same format as Fig.~\ref{fig:rixs_feII}.  }\label{fig:rixs_feIII}
\end{figure}

Similarly, we divide the RIXS spectra of the ferric complexes into three regions as shown in Fig.~\ref{fig:rixs_feIII}.
In the first region (0--1 eV), both the experimental (left panels) and theoretical spectra with different active space models (center and right panels) show a single band at 0 eV, similar to experiment.
In the second region (1--4 eV), the experimental spectrum of the chloride complex shows three representative bands at 1.83, 2.34, and 2.80 eV, while the thiolate complex exhibits a broader band with indistinct peaks in the range of 0.5--3.5 eV.
The minimal active space model yields two main bands at 2.58 and 3.42 eV for the chloride complex, and a slightly broader band in the range of 2--3.5 eV for the thiolate complex.
In the chloride complex, the larger active space model splits the two bands into three main bands at 2.46, 3.44, and 3.88 eV and two minor bands at 2.78 and 3.00 eV.
In the thiolate complex, in the larger active space the bands away from 0 eV are broadened, but still do not match the widths of the experimental bands.
Experimental measurements using magnetic circular dichroism (MCD) spectroscopy~\cite{gebhard1991single} show d-d transition bands between $0.9 \sim 1.39$ eV, while the ligand-to-metal charge transfer (LMCT) bands begin at $\sim 1.6$ eV and extend to $\sim 3.7$ eV for $\ce{[Fe^{III}(SDur)_4]^{1-}}$.
It is likely that the 
missing LMCT contributions in the theoretical active space lead to bands that are too narrow and are positively shifted, as well as the absence of bands and features in the high energy region ($>$ 4 eV).


\begin{figure}[ht!]
\center{\includegraphics[width=\columnwidth]{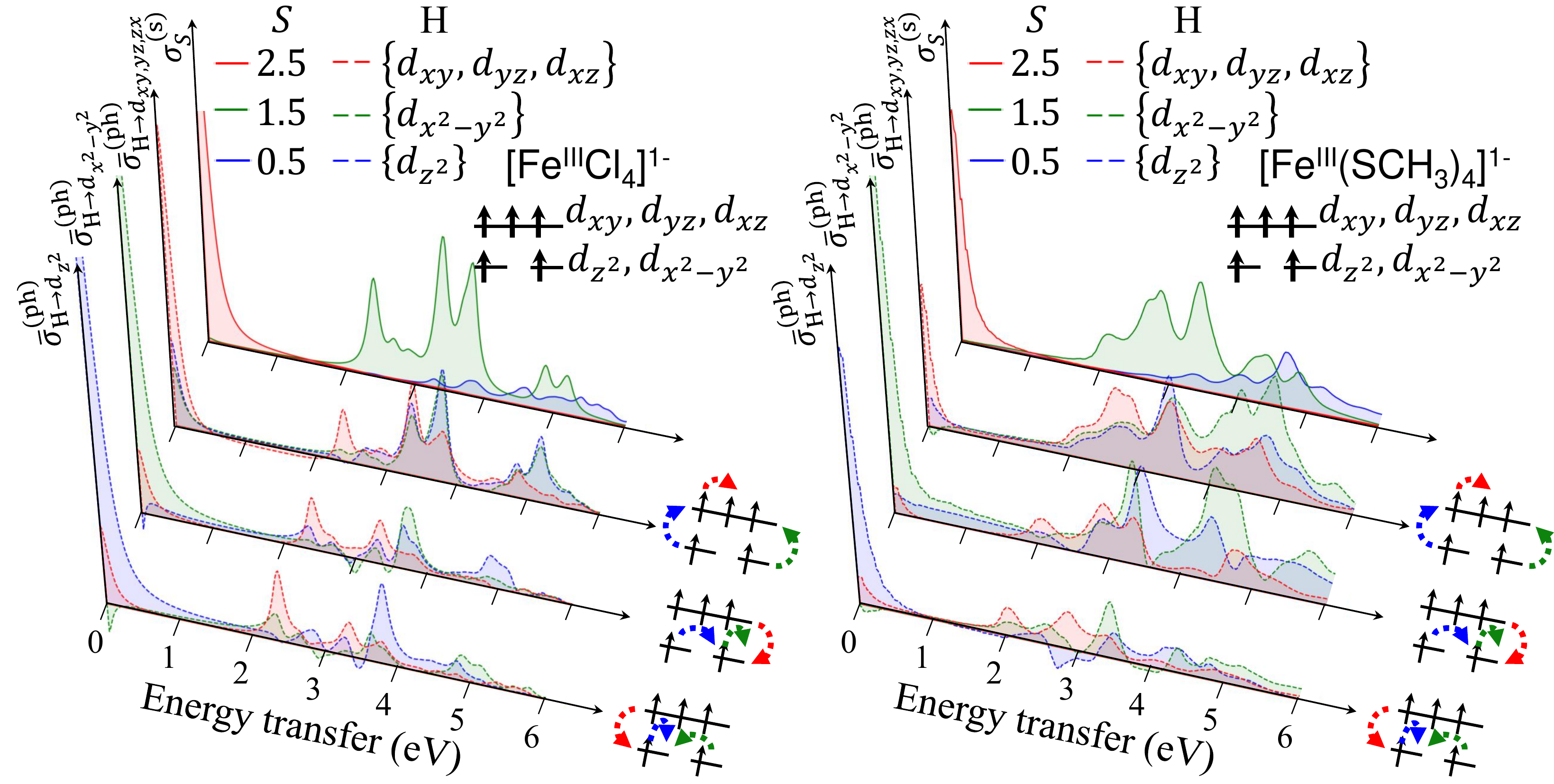}}
\caption{ Deconvolved theoretical RIXS spectra for $\ce{[Fe^{III}Cl_4]^{1-}}$ (left panels) and $\ce{[Fe^{III}(SCH_3)_4]^{1-}}$ (right panels), with the same format as Fig.~\ref{fig:feII_rixs_deconv}.
}\label{fig:feIII_rixs_deconv}
\end{figure}

Finally, we deconvolve the theoretical spectra of the larger active space model, using the same schemes as used in Fig.~\ref{fig:feII_rixs_deconv}.
Figure~\ref{fig:feIII_rixs_deconv} shows the deconvolved spectra for \ce{[Fe^{III}Cl_4]^{1-}} and \ce{[Fe^{III}(SCH_3)_4]^{1-}} in the left and right panels, respectively.
Like in the ferrous complex, the first region (0--1 eV) is dominated by spin-allowed transitions, while the second and third regions (1--6 eV) are characterized by spin-forbidden transitions of $\Delta S = 1, 2$.
We observe that 
in the ferric complexes, there is a larger contribution from the $\Delta S = 2$ transitions as compared to the ferrous complexes.
Furthermore, the other spectra
in the front and middle panels 
show the average particle-hole contributions from three sets of particles and holes, namely, $\{ d_{z^2} \}$, $\{ d_{x^2-y^2} \}$, and $\{ d_{xy}, d_{yz}, d_{xz} \}$.
In contrast to the ferrous complexes, all transitions to $d_{z^2}$, $d_{x^2-y^2}$, and $d_{xy, yz, xz}$ contribute to the total spectra across the entire energy range of 0--6 eV in the ferric complexes.
The low energy part of the band in the 2--3 eV region is dominated by spin-flip transitions of $d_{xy, yz, xz} \rightarrow d_{z^2}, d_{x^2-y^2}, d_{xy, yz, xz}$ character, while above this window there is a mixture of various particle-hole contributions.

\section{Conclusion} \label{sec:conclusions}

In this work, we presented a new ab initio technique to  compute L${}_{2,3}$-edge XAS and 2p3d RIXS spectra, based on the correction vector approach and a restricted active space ansatz.
We obtain good general agreement between our theoretical simulations and experimental spectra in a set of mononuclear tetrahedral ferrous and ferric iron complexes. 
Our results highlight the importance of selecting an appropriate active space for the spectroscopy, and the role of electron correlation between the metal and ligand electrons in determining certain spectral features. 
Improved simulations should incorporate additional orbitals into the active space treatment, for example, to treat LMCT/MLCT states; improve the active space through orbital optimization; and include dynamic correlation effects in the spectra.

The elimination of a sum-over-states computation in the correction vector formulation of XAS and RIXS spectra removes a major limitation in the simulation of spectra for multi-nuclear transition metal complexes. 
Simulations of larger iron-sulfur cluster X-ray spectra are currently underway in our group and will be presented elsewhere.

\section*{Acknowledgments}
SL was supported by the U.S. Department of Energy, Office of Science, via grant no. DE-SC0019374.
Additional support for SL was provided by the U.S. Department of Energy, Office of Science, National Quantum Information Science Research Centers, Quantum Systems Accelerator. GKC was supported by the US. Department of Energy, via grant no. DE-SC0023318.

\bibliographystyle{naturemag}
\bibliography{XAS_RIXS}

\end{document}